\documentclass[aps,twocolumn,showpacs]{revtex4}
\usepackage{amsmath,amssymb}
\usepackage{graphicx}

\newcommand{\ldangle}{\langle\!\langle}
\newcommand{\rdangle}{\rangle\!\rangle}

\begin{document}

\title{Bound state of a hole and a triplet spin 
in the $t_1$-$t_2$-$J_1$-$J_2$ model}

\author{Kazuhiro  Sano}
\affiliation{Department of Physics Engineering,  Mie University, Tsu, Mie 514-8507, Japan} 
\author{Ken'ichi Takano}
\affiliation{Toyota Technological Institute, Tenpaku-ku, Nagoya 468-8511, Japan}

\date{\today}

\begin{abstract}
We show that a hole and a  triplet spin form a bound state in a nearly half-filled band of the one- and two-dimensional $t_1$-$t_2$-$J_1$-$J_2$ models. 
Numerical calculation indicates that the bound state is a spatially small object and moves as a composite particle with spin 1 and charge  $+e$ in the spin-gapped background. 
Two bound states repulsively interact with each other in a short distance and move independently as long as they keep their distance. 
If a finite density of bound states behave as bosons, 
the system  undergoes the Bose-Einstein condensation 
which means a superconductivity with charge $+e$. 
\end{abstract} 

\maketitle

\section{Introduction}\label{intro}

To find an exotic mechanism of superconductivity is a fascinating and difficult challenge. 
A category of possible mechanisms is based on the hole motion in strongly correlated electrons with a nearly half-filled electron band.\cite{Kivelson,Trugman,Kane,White,Sano1,Takano-Sano} 
Then the mechanism of superconductivity is the Bose-Einstein condensation (BEC) of the holes. 
Although the mechanism was originally considered for explaining  cuprate superconductors, it may be appropriate for materials to be found in future. 
The hole is also closely related with metal-insulator transition and magnetism, since  background electron system is affected  by the hole motion. 
In spite of the importance, we have not arrived at a consensus on the precise role of the hole motion in superconductivity. 
This is because the motion of the strongly-correlated background electrons surrounding and forming the holes is a difficult quantum many-body problem. 
To overcome the difficulty and achieve a solid understanding of the hole motion, exhaustive studies by  numerical methods  for  finite-size systems are desirable.

\begin{figure}[b] 
\begin{center}\leavevmode
\includegraphics[width=0.33\linewidth]{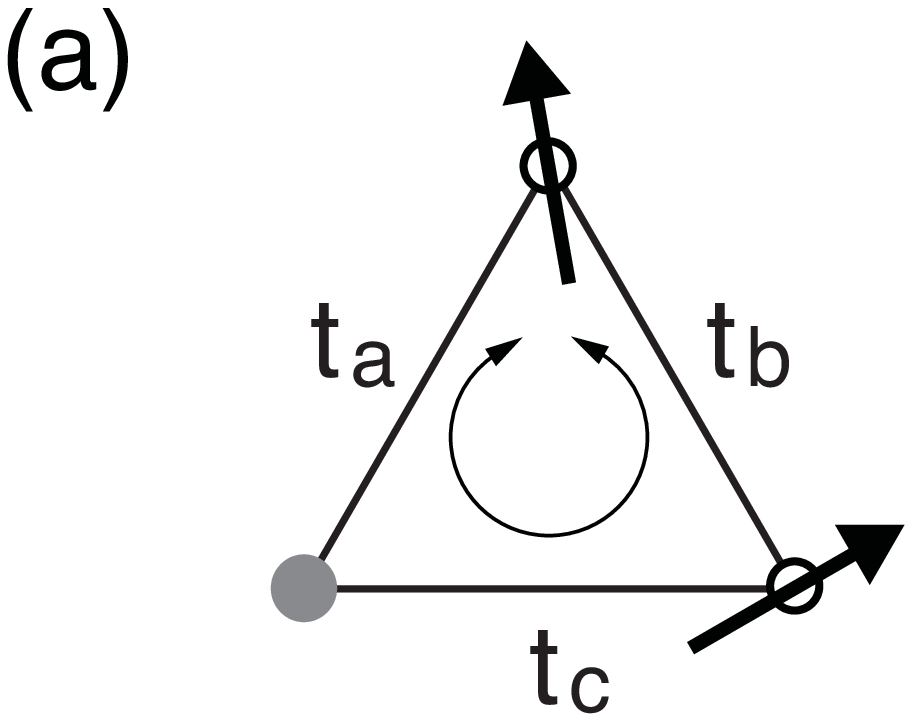}
\end{center}
\begin{center}\leavevmode
\includegraphics[width=0.75\linewidth]{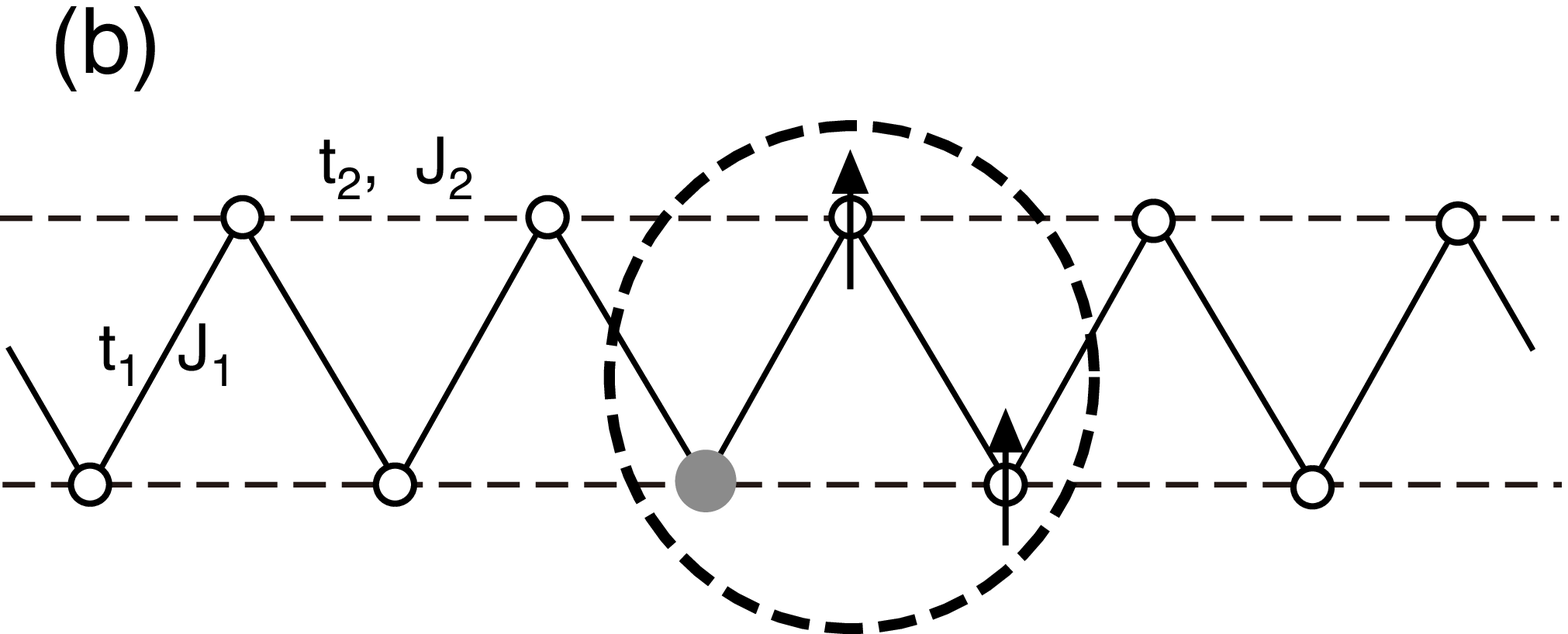}
\end{center}
\caption{
(a) Illustration for the triangle effect. 
The figure represents a typical triangle unit in nonbipartite lattices. 
The transfer integrals are $t_a$, $t_b$, and $t_c$ and infinite on-site repulsions are considered. 
An open circle is a site occupied by an electron and an arrow is the image of the spin. 
The site filled by gray is not occupied by an electron, meaning a hole. 
See text. 
(b) 1D $t_1$-$t_2$-$J_1$-$J_2$ model with a nearly half-filled band for $t_2<0$ and $J_1>0$. 
A typical snapshot configuration of a hole and a triplet spin pair in a bound state is indicated by the bold dashed loop. 
} 
\label{1d-lattice}
\end{figure}

In previous works,\cite{Sano1,Takano-Sano,Doi,tjdimer} 
we studied the hole motion  for  several strongly correlated electron systems in one-dimensional (1D) and two-dimensional (2D) systems by the numerical diagonalization. 
These systems include triangle units of three sites. 
We illustrate a typical triangle unit in Fig.~\ref{1d-lattice}(a). 
The triangle unit is described by the Hamiltonian consisting of three hopping terms with transfer integrals $t_a$, $t_b$, and $t_c$ under the restriction representing the infinite on-site repulsions. 
The ground state of this triangle unit is a singlet state for $t_a t_b t_c >0$, and is a triplet state for $t_a t_b t_c <0$. 
This is a consequence of the nonbipartite lattice structure connected by the three hopping terms. 
We call this the {\it triangle effect}. 
Based on the triangle effect, we particularly examined the 1D $t_1$-$t_2$-$J_1$-$J_2$ model, i.~e. the $t$-$J$ model in a zigzag chain, as shown in Fig.~\ref{1d-lattice}(b), when the number of electrons is less than that of lattice sites by just  one.\cite{Doi} 
We obtained the ground-state phase diagram in the space of parameters $t_2$ and $J_1$ in the case of $J_2 = (t_2/t_1)^2 J_1$. 
Then, we found a fairly large phase with total spin 1 in the region  of $t_2<0$ and $J_1>0$ of the phase diagram. 

By inspecting the numerically obtained ground-state wavefunction, we found the picture that the hole and a short triplet spin pair form a bound state to move in a singlet electronic background, as illustrated in Fig.~\ref{1d-lattice}(b). 
The triplet spin pair in the hole-spin bound state is from the spin degrees of freedom of the background electrons. 
In contrast, the charge degrees of freedom of the electrons are dead due to the infinite on-site repulsion except for the collective charge $+e$ of the hole. 
As the hole is transferred, the triplet spin pair changes into a singlet spin pair and another singlet spin pair close to the hole changes into a triplet spin pair.
By treating the hole-spin bound state as a free composite particle, 
we numerically obtained the dispersion relation approximately proportional to $k^2$ ($k$: the wave number) for long wavelengths. 
The effective transfer integral is then estimated as about $0.24|t_1|$.\cite{Doi} 
If a finite density of hole-spin bound states are stable and bosonic in more than one dimension, we expect that the BEC takes place at sufficiently low temperatures. 
Then the BEC is considered to be an exotic superconductivity with charge $+e$. 

In this paper, we investigate the hole motion and its influence to the 
surrounding electron spins in the 1D and 2D $t_1$-$t_2$-$J_1$-$J_2$ models by the numerical diagonalization method. 
We confirm that a hole and a triplet spin pair form a bound state in the ground-state phase with total spin $S=1$. 
In the 1D case, we reexamine the existence of the hole-spin bound state in detail. 
Further, we examine the two hole case by calculating low excitation energies and the density-density correlation function, and 
show that two hole-spin bound states exist stably and interact repulsively with each other. 
In the 2D case, we argue the existence of a similar hole-spin bound state by the numerical diagonalization method for small-size systems. 
We thus have a basis to argue the BEC of hole-spin bound states, i.e. a superconductivity. 

This paper is organized as follows. 
In Sec. \ref{hamiltonian}, we present and explain the Hamiltonian of the $t_1$-$t_2$-$J_1$-$J_2$ model. 
In Sec. \ref{onedim}, we confirm the stability of the hole-spin bound state in the 1D case by using the numerical diagonalization method. 
In Sec. \ref{twodim}, the 2D and quasi-1D cases are examined and the hole-spin bound state similar to that of the 1D case is argued. 
Section \ref{discussion} is devoted to summary and discussion. 

\section{ Hamiltonian}\label{hamiltonian}

We consider the 1D and 2D $t_1$-$t_2$-$J_1$-$J_2$ models. 
They are commonly represented by the Hamiltonian: 
\begin{align} 
H = &-t_{1}\sum_{\langle ij \rangle} \sum_{\sigma} 
(c_{i,\sigma}^{\dagger} c_{j,\sigma}+h.c.)  
\nonumber \\
&-t_{2}\sum_{\ldangle i'j' \rdangle} \sum_{\sigma} 
(c_{i',\sigma}^{\dagger} c_{j',\sigma}+h.c.) 
\nonumber \\
&+J_{1}\sum_{\langle ij \rangle} 
\left({\bf S}_{i} \cdot {\bf S}_{j}-\frac{1}{4}n_in_{j}\right)  
\nonumber \\
&+J_{2}\sum_{\ldangle i'j' \rdangle} 
\left({\bf S}_{i'} \cdot {\bf S}_{j'}-\frac{1}{4}n_{i'}n_{j'}\right) , 
\label{tJ-Hamil} 
\end{align}  
where $c^{\dagger}_{i,\sigma}$ is the creation operator of 
an electron with spin $\sigma$ at site $i$, 
${\bf S}_{i}$ = 
$\sum_{\sigma \sigma'} 
c_{i,\sigma}^{\dagger} ({\bf \tau})_{\sigma \sigma'} 
c_{i,\sigma'}$ with Pauli matrices ${\bf \tau}$, and 
$n_{i,\sigma}=c_{i,\sigma}^{\dagger}c_{i,\sigma}$. 
The summations with 
$\langle ij \rangle$ and $\ldangle i'j' \rdangle$ 
are taken over nearest neighbor (NN) and next nearest neighbor (NNN) pairs of sites, respectively. 
We take account of the infinite on-site repulsion by removing states with doubly occupied sites from the Hilbert space. 

The hopping energies between NN sites and between NNN sites are denoted by $t_1$ and $t_{2}$, respectively, and the corresponding exchange energies are denoted by $J_1$ and $J_{2}$, respectively. 
For our numerical examination to be of realistic amount, we need to reduce the number of parameters. 
We concentrate on the case of 
\begin{align} 
J_2=\left(\frac{t_2}{t_1}\right)^2J_1 
\label{condition}
\end{align} 
through this paper. 
This condition preserves the approximate equivalence between our model and the Hubbard model,  as long as $J_1$ and $J_2$ are not very large in comparison with $t_1$. 
In Eq.~(\ref{tJ-Hamil}), the sign of $t_1$ has no physical effect, 
while the system is sensitive to that of $t_2$. 
In what follows we set as $t_1=1$ without spoiling generality. 

We denote the total  number of sites as $N$, and the number of holes in the half-filled band as $N_{\rm h}$. 
Then the hole density  is given by $n_{\rm h} = N_{\rm h}/N$. 
We also denote the magnitude of the total spin as $S$. 
We numerically diagonalize the Hamiltonian (\ref{tJ-Hamil}) 
by using the standard Lanczos algorithm to calculate energies and correlation functions. 
The system size $N$ which we consider is maximally 23 for 1D system and 16 for 2D system.

\section{One-Dimensional case}\label{onedim}

\begin{figure}[t] 
\begin{center}\leavevmode
\includegraphics[width=0.9\linewidth]{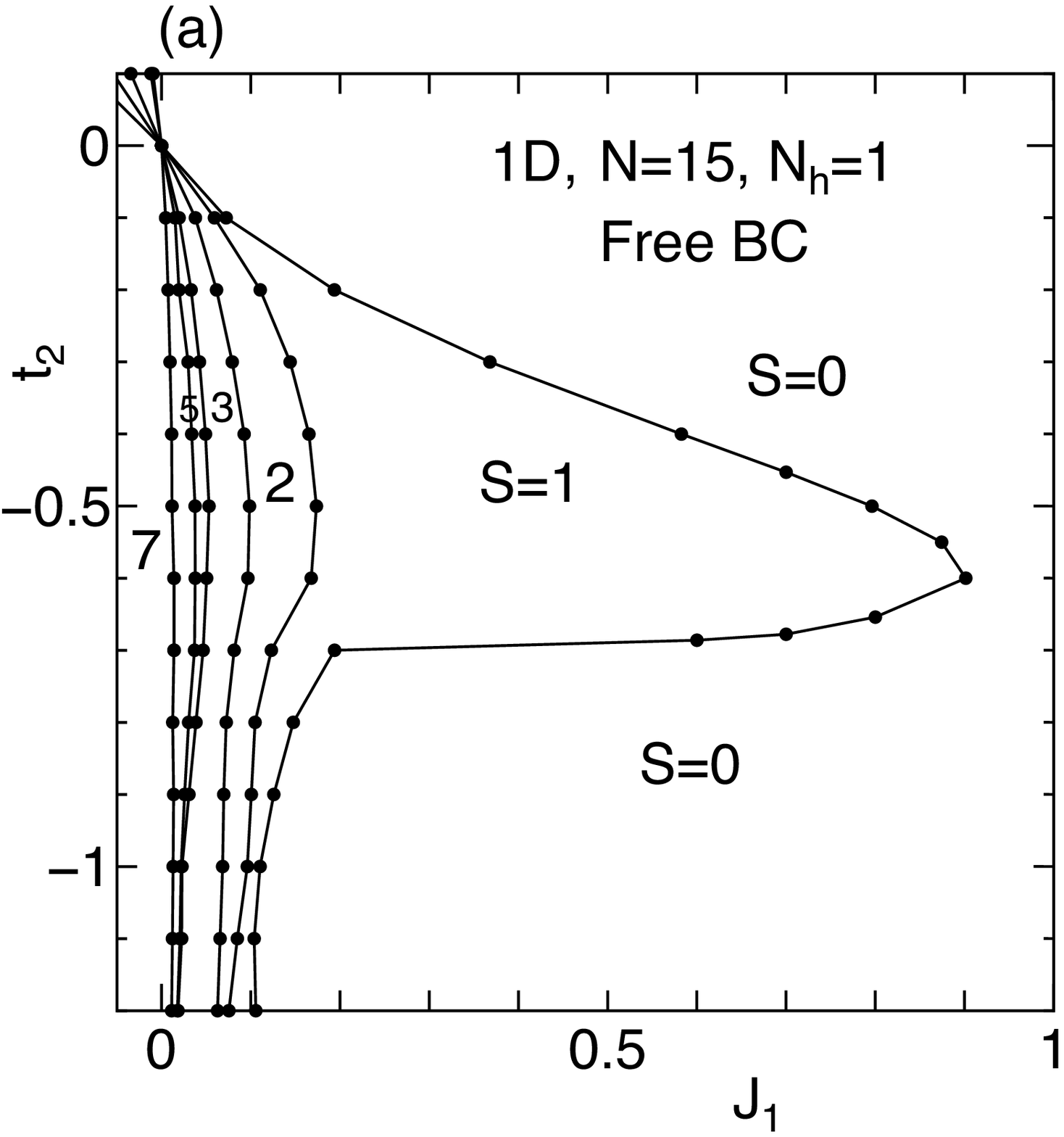}
\end{center}
\begin{center}\leavevmode
\includegraphics[width=0.9\linewidth]{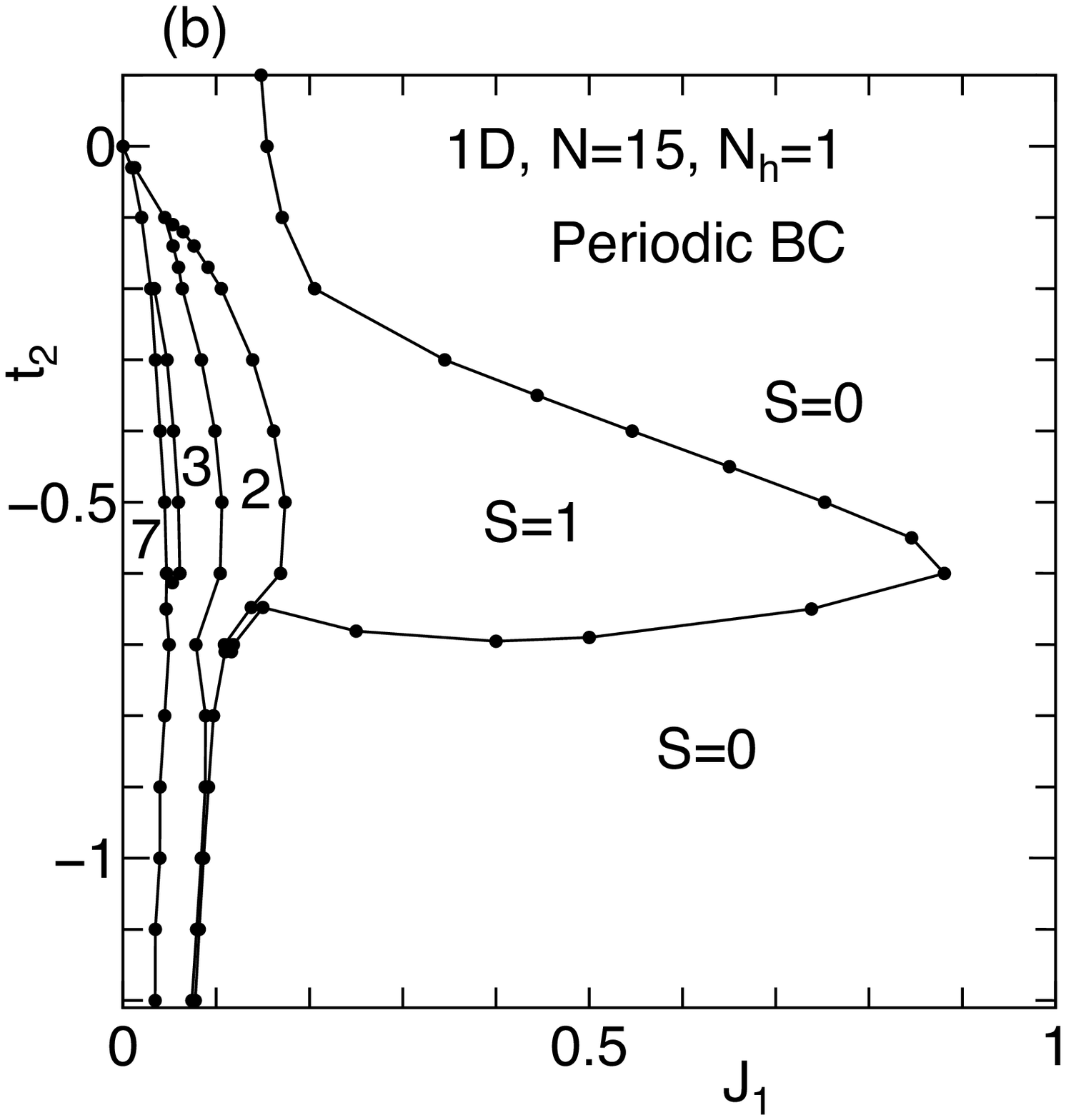}
\end{center}
\caption{Ground-state phase diagram for the one hole case 
in the 1D $t_1$-$t_2$-$J_1$-$J_2$ model with 
(a) the free boundary condition and 
(b) the periodic boundary condition.} 
\label{phase-d1}
\end{figure}

\begin{figure}[h] 
\begin{center}\leavevmode
\includegraphics[width=0.9\linewidth]{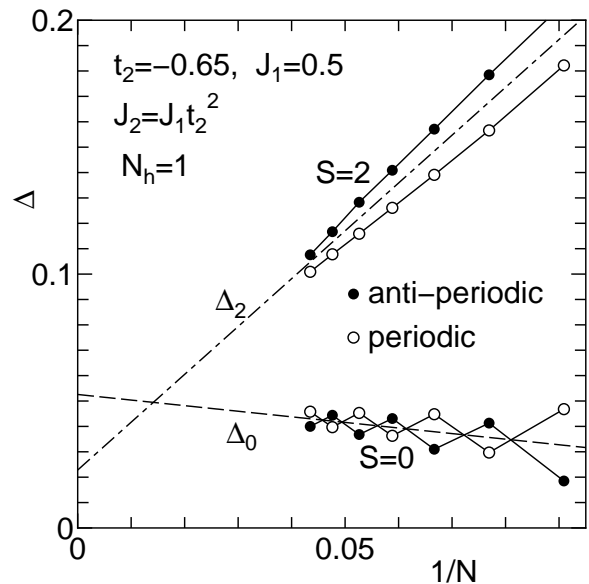}
\end{center}
\caption{Size dependence of the excitation energies from  the ground state with $S=1$ to the lowest  states with $S=0$ and $S=2$ for $N=11,13,15,17,19,21$, and $23$. 
The open circle and solid circle stand  results of the periodic boundary 
 condition and the antiperiodic boundary condition, respectively.}
\label{gap}
\end{figure}

The 1D $t_1$-$t_2$-$J_1$-$J_2$ model describes the spin chain shown in Fig.~\ref{1d-lattice}(b). 
We have examined this model by numerically diagonalizing the Hamiltonian (\ref{tJ-Hamil}) up to $N=13$ with the free boundary condition.\cite{Doi} 
Then we have obtained the ground-state phase diagrams of the one-hole case ($N_{\rm h}=1$). 
Each phase diagram has a relatively large phase of $S=1$ penetrating into an extended singlet phase. 
By inspecting the $S=1$ ground state, we have proposed that the hole and 
a triplet spin pair form the bound state in the ground state, as illustrated in Fig.~\ref{1d-lattice}(b). 

To ensure the existence of the $S=1$ phase, we examine the case of $N=15$ by using both the free and periodic boundary conditions. 
The resultant phase diagrams by the numerical diagonalization are shown in Figs.~\ref{phase-d1}(a) and (b), respectively. 
We observe that the shape and size of the $S=1$ region for $N=15$ in Fig.~\ref{phase-d1}(a) are almost the same as those for $N=13$.\cite{Doi} 
Also the comparison of Fig.~\ref{phase-d1}(a) to Fig.~\ref{phase-d1}(b) shows that the region of $S=1$ is almost irrespective of the boundary conditions. 
Since the finite size effect is small, the $S=1$ phase including the bound state of the hole and the 
triplet spin is expected to survive in the thermodynamic limit. 

\begin{figure}[t] 
\begin{center}\leavevmode
\includegraphics[width=0.7\linewidth]{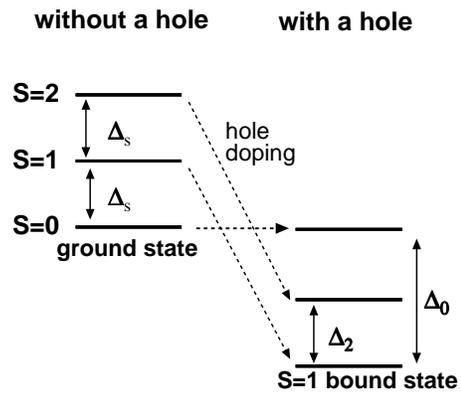}
\end{center}
\caption{Schematic diagram of energy levels of  systems with and without a
 hole.}
\label{e-level}
\end{figure}

The stability of the hole-spin bound state in the $S=1$ phase is reflected in typical excitation energies. 
We calculate the lowest excitation energies $\Delta_{0}$ and $\Delta_{2}$ to the $S=0$ and $S=2$ states, respectively, by the numerical diagonalization. 
We employ both the periodic and antiperiodic boundary conditions to reduce finite size effects. 
Figure \ref{gap} shows the calculated $\Delta_0$ and $\Delta_2$ as functions of $N$. 
By extrapolating the values for the finite systems, we have $\Delta_{0} \simeq 0.052$ and $\Delta_{2} \simeq 0.024$ in the thermodynamic limit. 
We explain these excitations by using the schematic energy-level diagram for systems with and without hole in Fig.~\ref{e-level}; the system without hole is a pure spin model, i.~e. the $J_1$-$J_2$ model. 
(i) The excitation energy $\Delta_{2}$ is interpreted as the spin gap of spins except for the localized hole-spin bound state, and corresponds to  the spin gap $\Delta_{s}$ of the pure spin system. 
The energy correspondences are shown as two dotted arrows connecting the $S=1$ levels and connecting the $S=2$ levels in Fig.~\ref{e-level}. 
If the bound state with $S=1$ is sufficiently localized, the spin excitation with $S=1$ in the background is almost independent of the hole motion. 
In fact, the present value $\Delta_{2} \simeq 0.024$ is consistently close to the spin gap $\Delta_{s} \sim 0.03$ 
of the $J_1$-$J_2$ model at $J_2/J_1=0.65^2 \simeq 0.42$\cite{Tonegawa}. 
(ii) The finite gap $\Delta_{0}$ means the stability of the hole-spin bound state, where  the value $\Delta_{0}+\Delta_{s}$ corresponds to the formation energy of the hole-spin bound state as shown in Fig.~\ref{e-level}. 
If the hole-spin bound state vanishes, 
it resolves into an isolated hole and a 
triplet spin pair, and then a triplet pair changes to a singlet pair to be absorbed into the background electrons. 

\begin{figure}[t] 
\begin{center}\leavevmode
\includegraphics[width=0.9\linewidth]{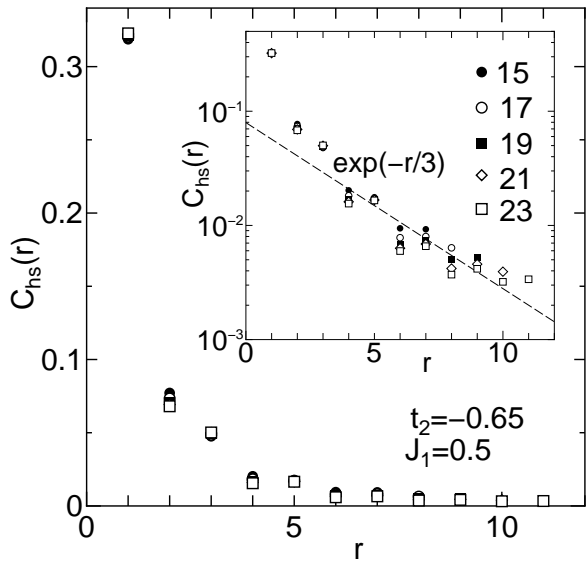}
\end{center}
\caption{Correlation  function $C_{\rm hs}(r)$ for $N=15,17,19,21$, and $23$ systems. Inset  shows the semi-log plot of $C_{\rm hs}(r)$.}
\label{spind}
\end{figure}

To see the relative distance between the hole and the 
triplet spin in the hole-spin bound state, we introduce the hole-spin correlation function: 
\begin{align} 
C_{\rm hs}(r) = \langle \, n_{\rm h}(i) \, s(i+r) \, \rangle ,
\end{align} 
where $n_{\rm h}(i) \equiv 1-n_{i,\uparrow}-n_{i,\downarrow}$ 
and $s(i) \equiv \frac{1}{2}(n_{i,\uparrow}-n_{i,\downarrow})$ are, respectively, the hole number and the $z$-component of the spin at site $i$. 
$C_{\rm hs}(r)$ characterizes the distance of the spin density measured from the position of the hole. 
By the numerical diagonalization, we obtained 
$C_{\rm hs}(r)$ for the cases of $N=15,17,19,21$, and $23$ in the subspace of total spin $S_z=1$. 
We only display the result for the antiperiodic boundary condition in Fig.~\ref{spind}, since the result for the  periodic boundary condition is almost the same. 
As is seen, the finite size effect for $C_{\rm hs}(r)$ is very small. 
We also show the semi-log plot of $C_{\rm hs}(r)$ against the distance $r$ in the inset. 
By the fitting, we find that $C_{\rm hs}(r)$ decays with an exponential factor $\exp (-r/3)$ and the spin density is concentrated near the hole. 
Thus the hole-spin bound state behaves as a compact composite particle. 

\begin{figure}[t] 
\begin{center}\leavevmode
\includegraphics[width=0.9\linewidth]{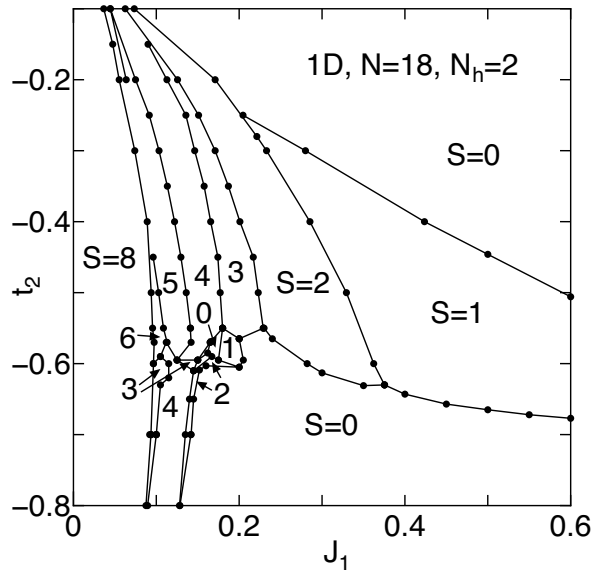}
\end{center}
\caption{Ground-state phase diagram for the two hole case 
in the 1D $t_1$-$t_2$-$J_1$-$J_2$ model 
with the antiperiodic boundary condition.}
\label{souzu}
\end{figure}

Next, we examine the two-hole case ($N_{\rm h} = 2$) by the numerical diagonalization. 
In  Fig.~\ref{souzu}, we show the  phase diagram for the system of $N = 18$ and $N_{\rm h} = 2$.
There is a phase of $S=2$ including a point of $(J_1, t_2) = (0.3, -0.5)$.
This $S=2$ phase is smaller than the $S=1$ phase of the one-hole case. 
However, the energies of the lowest $S=1$ and $S=2$ states are nearly degenerate in the $S=1$ and $S=2$ phases which share a common boundary. 
Hence, it is possible that the $S=2$ phase is large comparable to that of $S=1$ for the one-hole case  in the thermodynamic limit. 

\begin{figure}[b] 
\begin{center}\leavevmode
\includegraphics[width=0.9\linewidth]{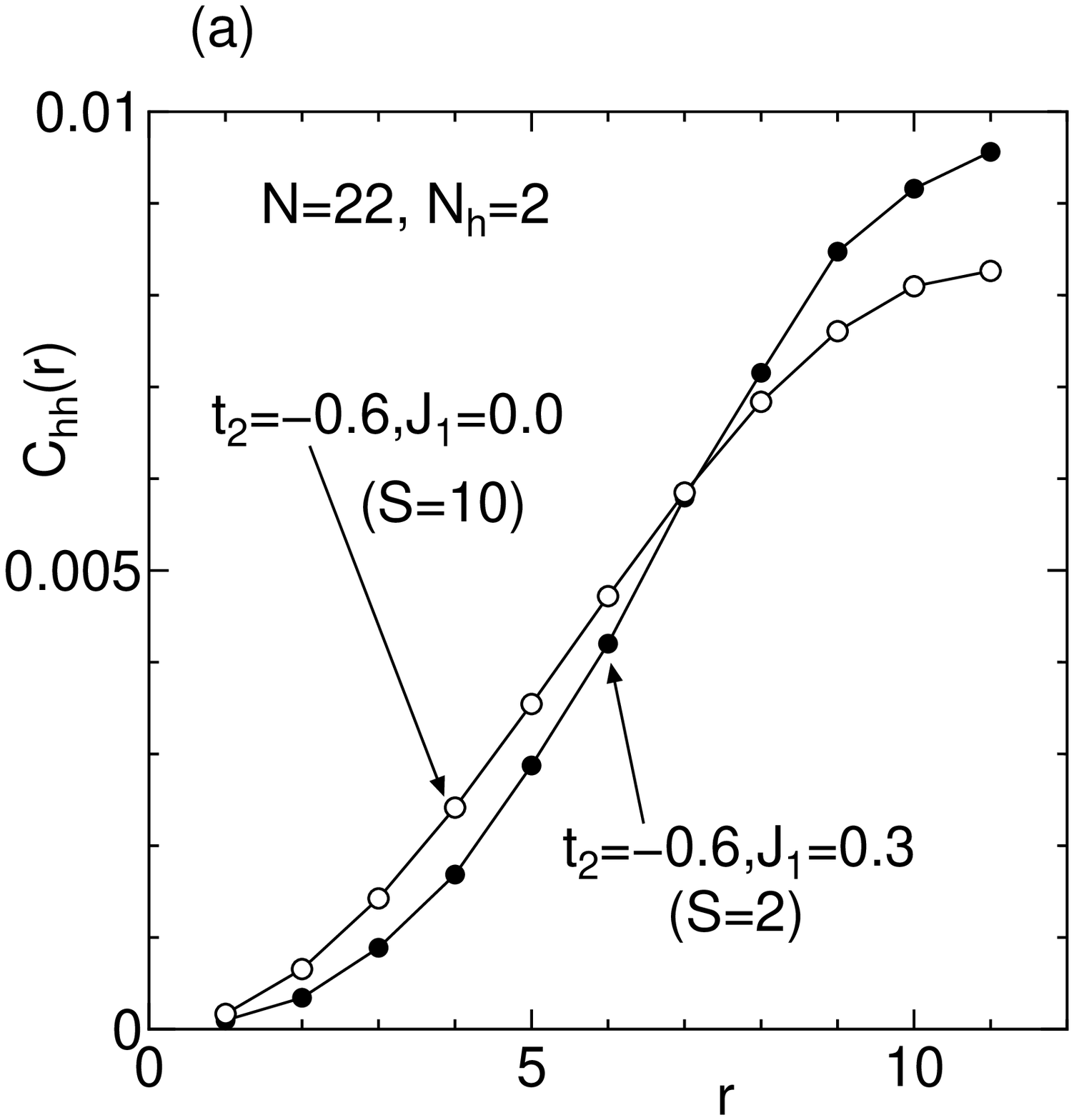}
\end{center}
\begin{center}\leavevmode
\includegraphics[width=0.9\linewidth]{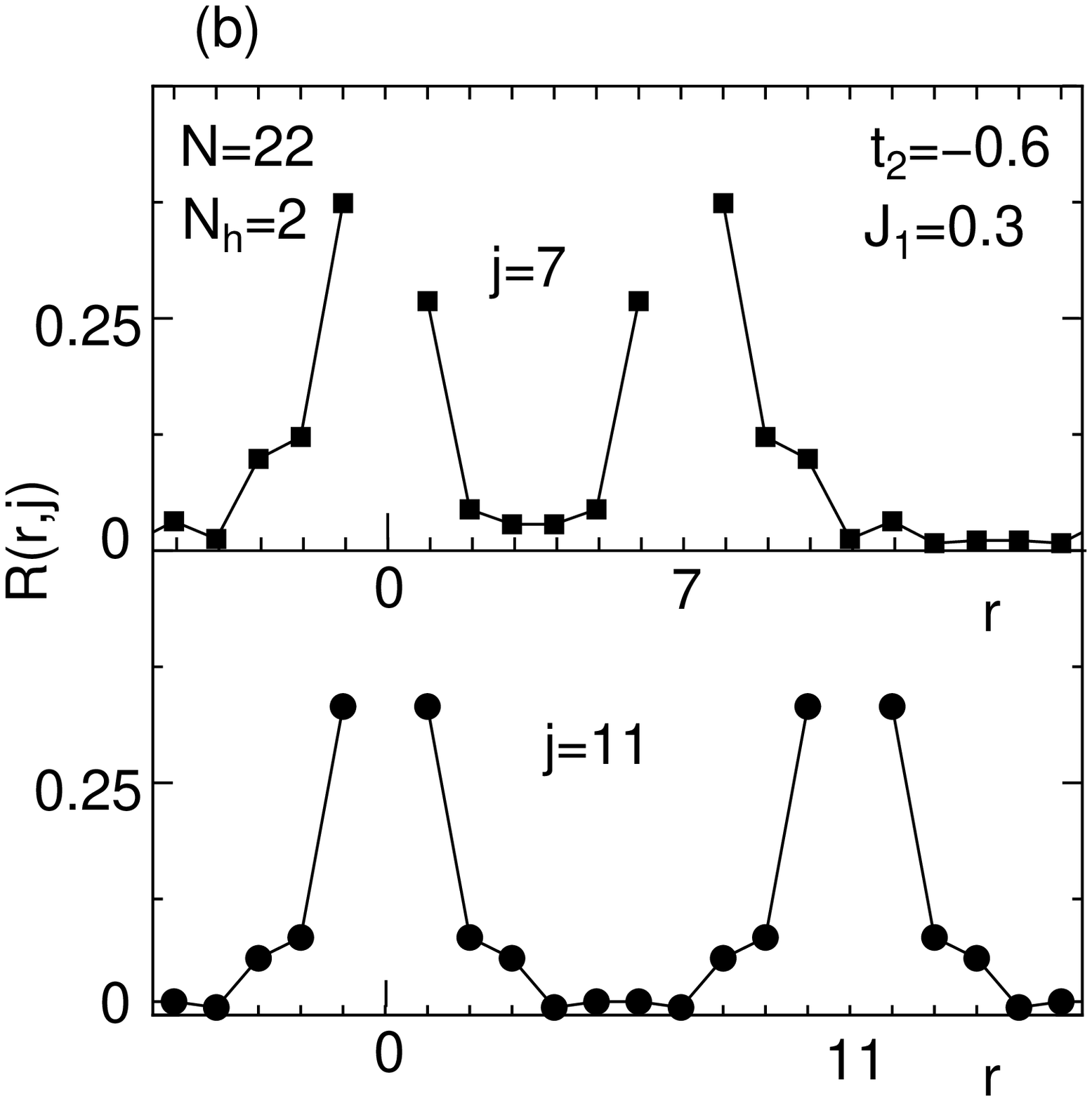}
\end{center}
\caption{(a) Density-density correlation function $C_{\rm hh}(r)$ between two holes for $N=22$ system with the antiperiodic boundary condition.
(b) Spin-density function $R(r,j)$ for the same system. 
$j$ is the distance between the two holes.
}
\label{CDW}
\end{figure}

If two hole-spin bound states are formed in the $S=2$ phase, the interaction between them comes from spin fluctuations since charge fluctuations are suppressed in the unmoved background electrons. 
Spin fluctuations may propagate as spin waves to induce the interaction between 
the hole-spin bound states. 
The formation of the spin gap suppresses spin waves with long wavelengths and confines the interaction in short range. 
To estimate the interaction range, we calculate the density-density correlation function $C_{\rm hh}(r)$ between two holes: 
\begin{align} 
C_{\rm hh}(r) = 
\langle \, n_{\rm h}(i) \, n_{\rm h}(i+r) \, \rangle . 
\end{align} 
Figure \ref{CDW}(a) shows $C_{\rm hh}(r)$ for $N=22$ and 
$N_{\rm h}=2$  at $(J_1, t_2) = (0.3, -0.60)$, and  also 
at $(J_1, t_2) = (0.0, -0.60)$ as a reference. 
We see that the two holes in the state of $(J_1, t_2)$ = $(0.3, -0.60)$ avoid each other for short distances ($r<7$) more than those in the reference state. 
In other words, the  interaction is more repulsive than  that of the reference state. 
In the reference state, holes  behave as noninteracting fermions, since the ground  state  is the fully polarized ferromagnetic state ($S=10$) and the exchange interactions are completely zero ($J_1=J_2=0.0$). 
Thus,  the hole motion at $(J_1, t_2) = (0.3, -0.60)$ is  really  repulsive. 

To examine the shape of the hole-spin bound state in the above system, we examine normalized spin density function defined by 
\begin{align} 
R(r,j)= \frac{\langle \, n_{\rm h}(i)n_{\rm h}(i+j)s(i+r) \, \rangle}{\langle \, n_{\rm h}(i)n_{\rm h}(i+j) \, \rangle} . 
\end{align} 
This represents the spin-density profiles of the two hole-spin bound states in the subspace of $S_z=2$, when one hole is at the site $i$ and the other is $i+j$. 
We show $R(r,7)$ and $R(r,11)$ for $N=22$, respectively, in the upper and lower panels of Fig. \ref{CDW}(b). 
In the upper panel where two holes are close, the spin density around a hole decreases in the side closer to the other hole. 
The result suggests that the two hole-spin bound states  avoid each other. 

The above results show that the two hole-spin bound states are stable and move almost independently by avoiding each other with a repulsive interaction for $N=22$. 
For the case of $N_{\rm h} \ge 3$, 
Sano and Ono have carried out the numerical diagonalization.\cite{Sano-ono}  
They found a partial ferromagnetic phase where the magnetization becomes weak when $n_h$ closes to the unity.
We interpret that the ferromagnetism is from incompletely formed hole-spin bound states due to the small system sizes.
We have not carried out calculations for systems with a finite density of hole-spin bound states due to the limitation of the numerical diagonalization method. 
We expect that independent hole-spin bound states are formed 
even for large systems including a finite density of holes. 

Nishimoto et al. studied the zigzag  Hubbard chain with strong on-site repulsion by the DMRG method\cite{Nishimoto}. 
They found an anomalous ground state with strong ferromagnetic fluctuation near the half filling at $t_2 \sim -1.0$. 
We suppose that the ground state of the $t_1$-$t_2$-$J_1$-$J_2$ model including a finite density of hole-spin bound states near the half-filling continues to the ground state of the Hubbard model.

\section{Two-Dimensional case}\label{twodim}

\begin{figure}[b] 
\begin{center}\leavevmode
\includegraphics[width=0.7\linewidth]{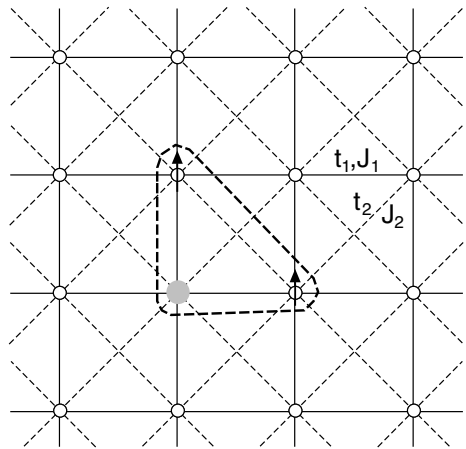}
\end{center}
\caption{
2D $t_1$-$t_2$-$J_1$-$J_2$ model with a nearly half-filled band for $t_2<0$ and $J_1>0$. 
Each open circle represents a site occupied by an electron, and the gray circle does a hole. 
We indicate a typical configuration of a hole-spin bound state consisting of the hole and a triplet spin pair by the loop of the bold dashed line, where the up-arrows mean electronic up-spins.
}
\label{latt-2d}
\end{figure}

\begin{figure}[!] 
\begin{center}\leavevmode
\includegraphics[width=0.9\linewidth]{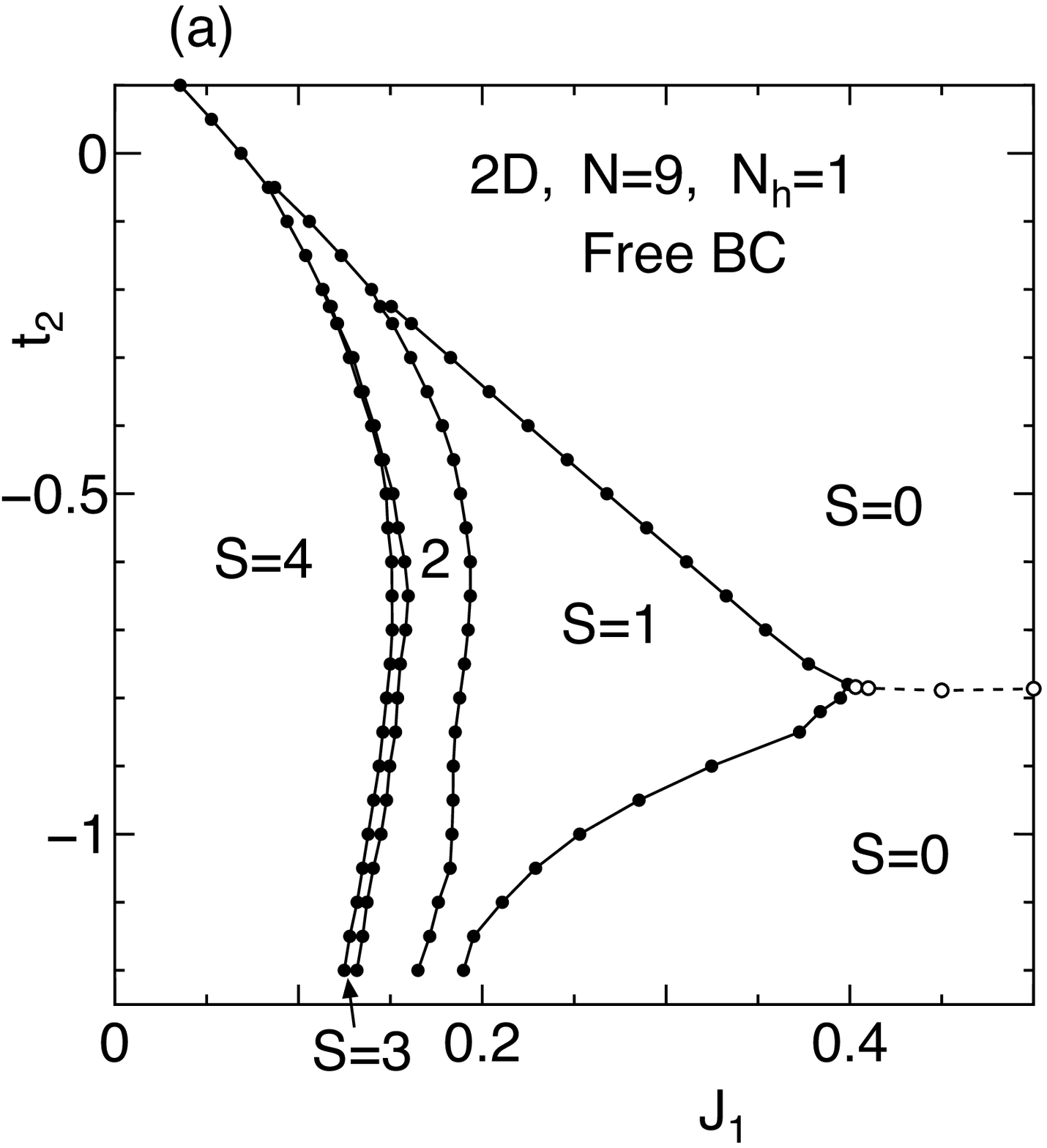}
\end{center}
\begin{center}\leavevmode
\includegraphics[width=0.9\linewidth]{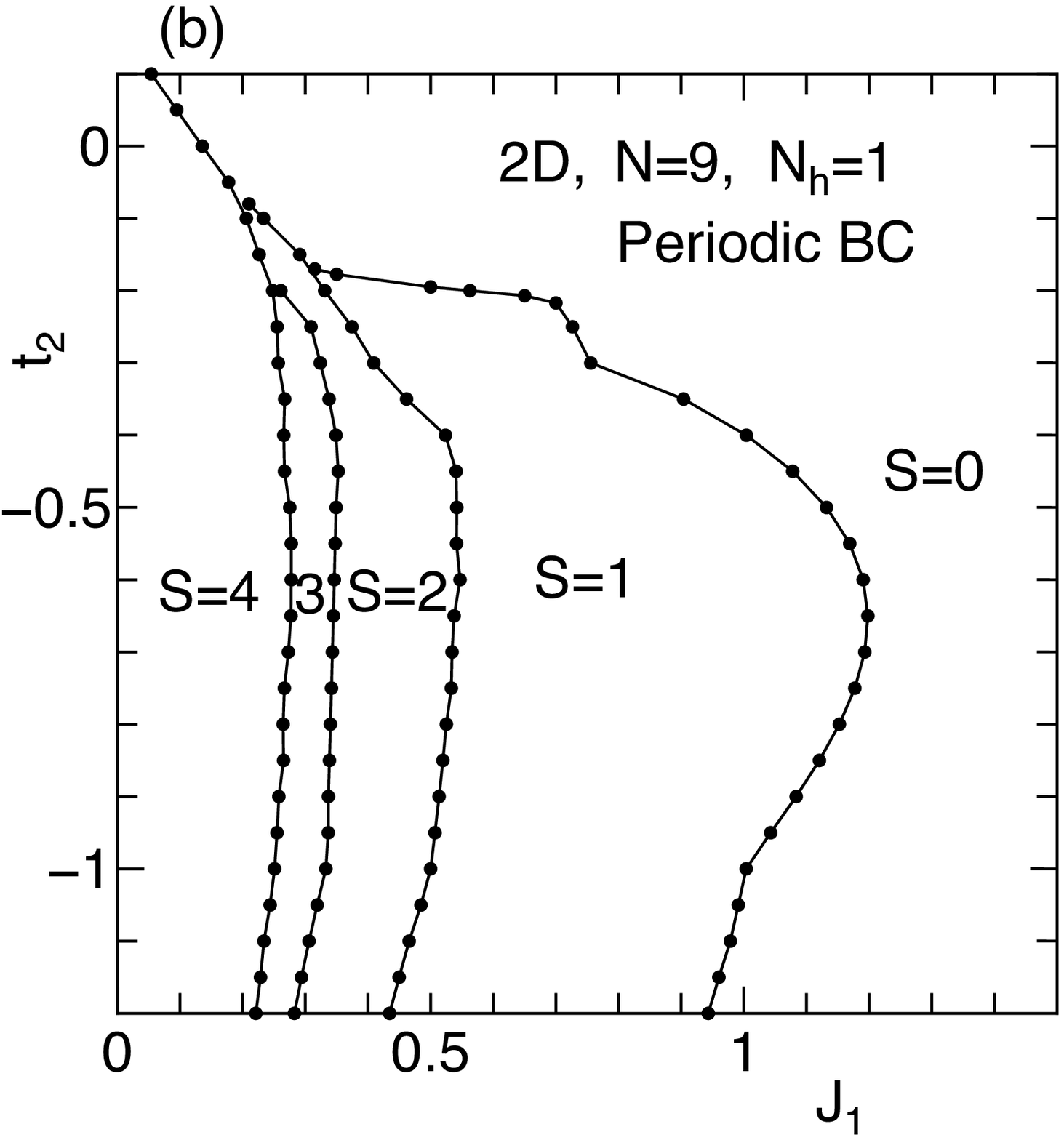}
\end{center}
\caption{Ground-state phase diagram for $3 \times 3$ lattice system in  the 2D $t_1$-$t_2$-$J_1$-$J_2$ model with 
(a) the free boundary condition and 
(b) the periodic boundary condition. 
The phase boundary between the $S=0$ phases are specially shown by dotted lines with open circles as calculated points; see text. 
}
\label{x2d-n9}
\end{figure}

\begin{figure}[t] 
\begin{center}\leavevmode
\includegraphics[width=0.9\linewidth]{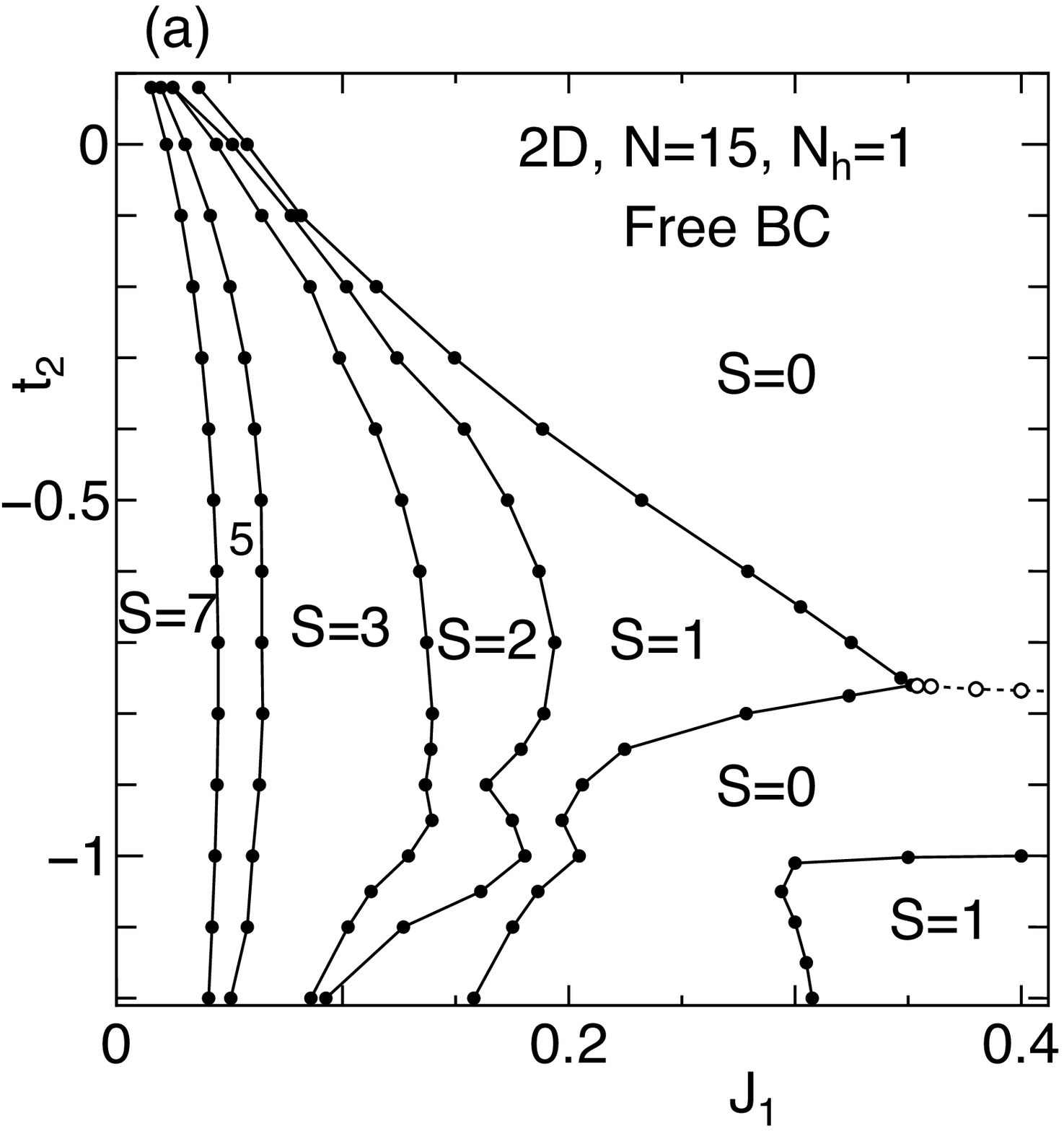}
\end{center}
\begin{center}\leavevmode
\includegraphics[width=0.9\linewidth]{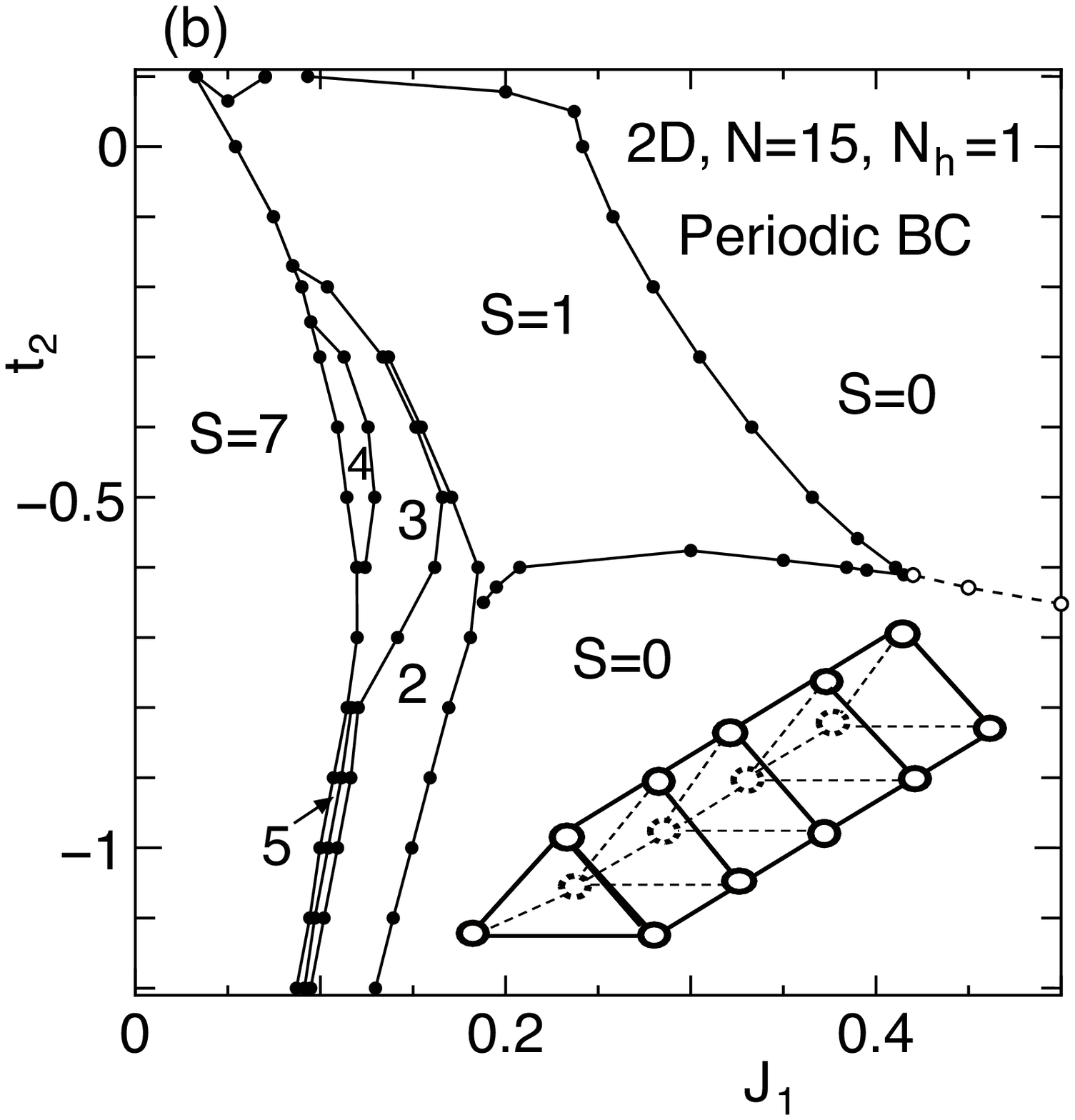}
\end{center}
\caption{Ground-state phase diagram for  the $3 \times 5$ lattice system with 
(a) the free boundary condition and 
(b) the periodic boundary condition. 
Inset of (b): The lattice structure under the periodic boundary condition. 
The lines representing $t_2$ and $J_2$ are abbreviated.}
\label{x2d-n15}
\end{figure}

\begin{figure}[t] 
\begin{center}\leavevmode
\includegraphics[width=0.9\linewidth]{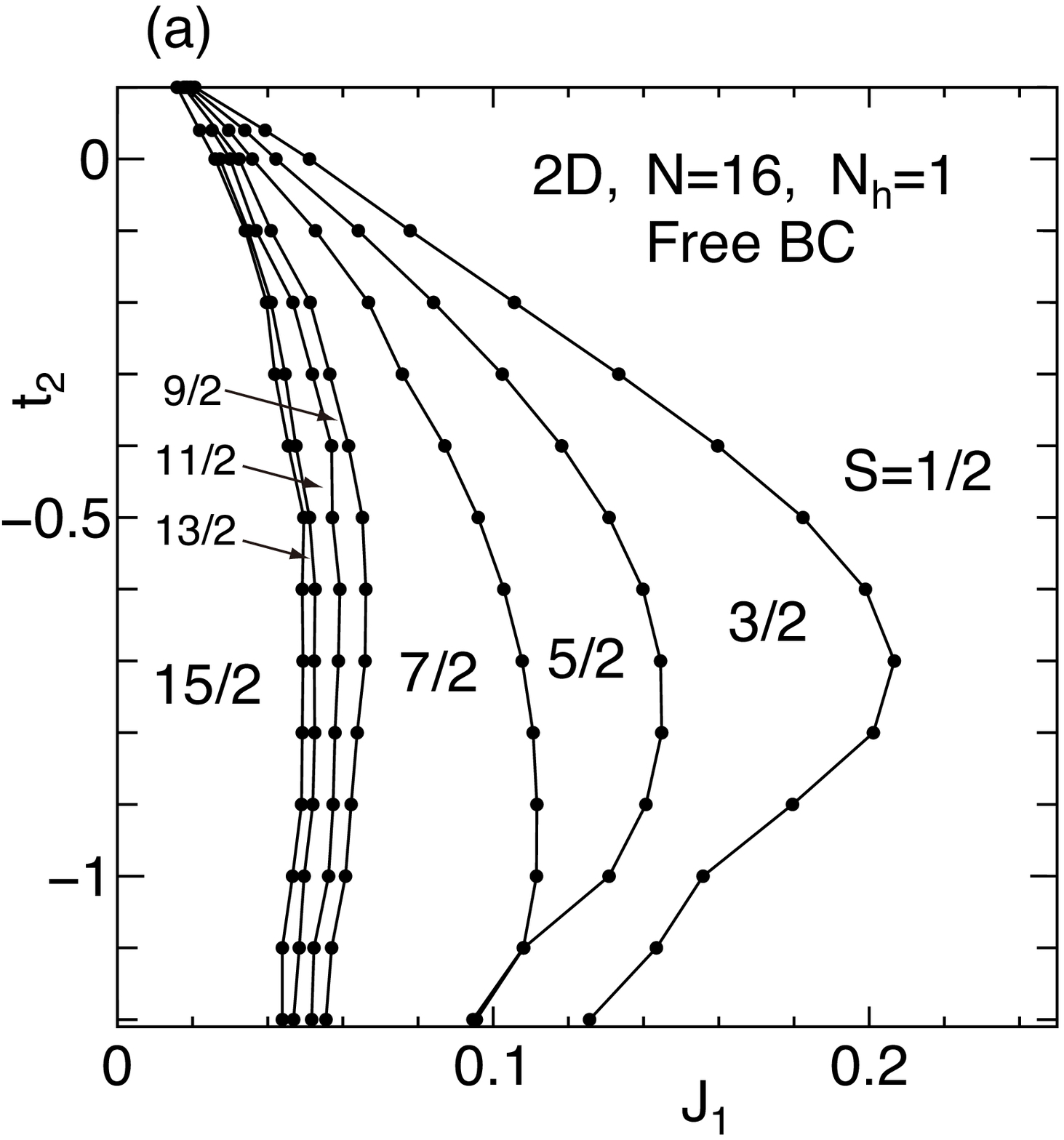}
\end{center}
\begin{center}\leavevmode
\includegraphics[width=0.9\linewidth]{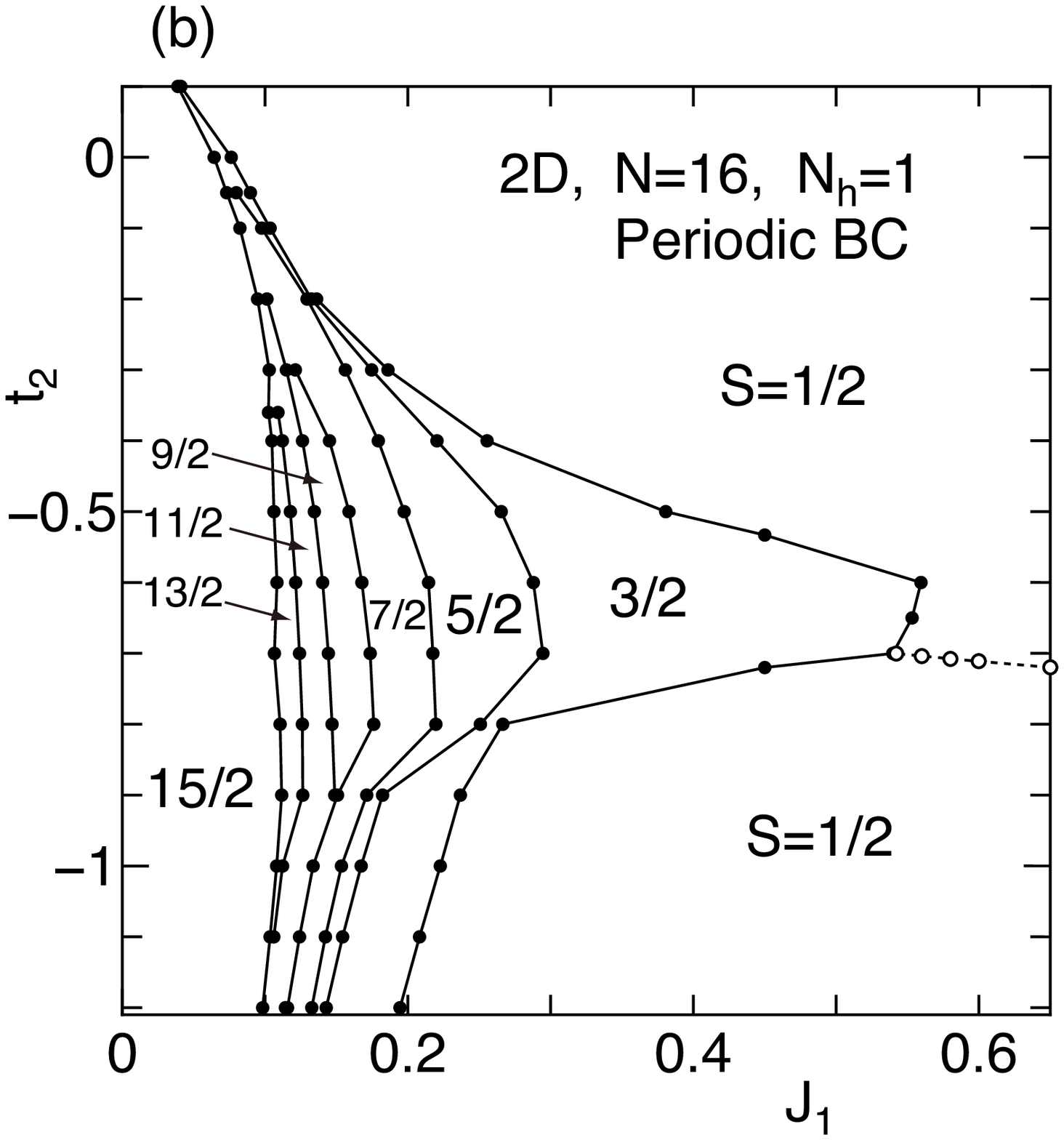}
\end{center}
\caption{Ground-state phase diagram for  the $4 \times 4$ lattice
 system with 
(a) the free boundary condition and 
(b) the periodic boundary condition.}
\label{x2d-n16}
\end{figure}

The 2D version of the $t_1$-$t_2$-$J_1$-$J_2$ model is shown in Fig.~\ref{latt-2d}. 
The hole-spin bound state consisting of a hole and a triplet spin pair
 similar to that for the 1D system is illustrated in this figure. 
We examine the $3 \times 3$, $3 \times 5$, and $4 \times 4$ lattice systems with both the free and periodic boundary conditions by the numerical diagonalization. 
For these systems, we obtained the ground-state phase diagrams in the $J_1$-$t_2$ plane as shown in Figs. \ref{x2d-n9}, \ref{x2d-n15}, and \ref{x2d-n16}. 
In what follows, we analyze the existence of the hole-spin bound state based on the phase diagrams. 

The case of the one hole and even number of electrons is realized in the $3 \times 3$ and $3 \times 5$ lattice systems. 
The phase diagrams of the systems with the free boundary condition are shown in Figs.~\ref{x2d-n9}(a) and \ref{x2d-n15}(a). 
At a glance, large $S=1$ phases are seen around $(J_1, t_2)$ = $(0.3, -0.6)$  in both the phase diagrams. 
The profiles of the phases are similar to that of the $S=1$ phase in Fig.~\ref{phase-d1}(a) for the 1D $t_1$-$t_2$-$J_1$-$J_2$ model. 
Hence we expect that the $S=1$ phase includes the hole-spin bound state.  
Another $S=1$ phase for $J_1 \gtrsim 0.3$ and $t_2 \lesssim -1.0$ in Fig.~\ref{x2d-n15}(a) is considered to be a finite size effect of the $3 \times 5$ lattice system. 

The bound-state phase is also seen for the periodic boundary condition. 
The phase diagram of the $3 \times 3$ lattice system for $t_1<0$ with the periodic boundary condition is shown in Fig.~\ref{x2d-n9}(b). 
The $S=1$ phase exists with an area larger than that for the free boundary condition. 
We consider that the different size is due to extra frustration from the periodic boundary condition. 
The phase diagrams of the $3 \times 5$ lattice system  for $t_1<0$ with the periodic boundary condition is shown in Fig.~\ref{x2d-n15}(b). 
Because of the periodicity, the lattice structure is of a tube where triangle units are stacked as shown in the inset of Fig.~\ref{x2d-n15}(b). 
We know that a hole motion in a triangle attract spin 1 so that the ground state with $S=1$ may be preferable even for $t_2 \sim 0$. 
It is also known that a spin tube without hole has a ground state with spin gap even for $J_2=0$.\cite{Okunishi,t-tube} 
These effects make the problem complicated near a regime around $t_2=0$, but does not deny the existence of the hole-spin bound state in a regime of the $S=1$ phase with intermediate value of $t_2$. 
Anyway, the different shape of the $S=1$ phase of the $3 \times 5$ lattice system from that of the $3 \times 3$ lattice system means that the effect of the periodic boundary effect is too large to infer the profile in the limit of large system size. 
In contrast, the phase shapes of the $S=1$ phases with the free boundary condition are similar to each other and encourage to consider the large system size limit. 

Here we consider the hole motion by beginning from the large $J_1$ limit; this also means the large $J_2$ limit owing to Eq.~(\ref{condition}). 
In this limit, the hole does not move and hence the system reduces to a frustrated spin system,  i.~e. the $J_1$-$J_2$ model, with a static defect at the hole site. 
The ground state of the $J_1$-$J_2$ model is a Neel state for $J_2/J_1 \lesssim 0.4$ and a collinear state for $J_2/J_1 \gtrsim 0.6$. 
For $J_2/J_1 \sim 0.5$, it is a plausible argument that there is a disordered phase with a spin-gapped ground state.\cite{Dagotto,Sorella,TKOS}
Now we reduce $J_1$ to be finite. 
Then the hole moves by the transfer terms with $t_1$ and $t_2$, and interacts with surrounding electrons. 
Since the hole motion is considered to destroy the Neel and collinear  spin orders, the spin-gap phase becomes more stable. 
Hence the spin-gap phase becomes wider than that in the $J_1$-$J_2$ spin system; or else another spin-gap phase may newly appear owing to the hole motion. 
The hole-spin bound state can survive in such a  background spin system with spin-gap. 
Unfortunately, the $3 \times 3$ and $3 \times 5$ systems in the present calculation are too small to directly detect the spin gap. 
However, the $S=1$ phase in the small systems is expected to survive even in the large system-size limit, by considering the above physical explanation of the destruction of the Neel and collinear orders by the hole motion. 
Further, in the 1D system, we have shown more plausibly the existence of the hole-spin bound state by the numerical diagonalization with relatively large system size. 
This suggests the existence of the hole-spin bound state in the 2D system which is in the same physical situation as that for the 1D system; 
i.~e. the triangle effect (Fig.~\ref{1d-lattice}(a)) commonly works 
in the gapped singlet backgrounds. 

For large $J_1$, we see phase boundaries between two $S=0$ phases in Figs. \ref{x2d-n9}(a), \ref{x2d-n15}(a), and \ref{x2d-n15}(b); 
in each figure, the phase boundary is drawn by a dotted line and calculated points are indicated by open circles. 
The level crossing is the change between the Neel ground state for small $J_2/J_1$ and the collinear ground state for large $J_2/J_1$.\cite{Dagotto,Sano-Doi-Takano} 
The phase boundary is terminated by the $S=1$ phase, which has a sharp corner at the terminal point. 
We see a single $S=0$ phase and do not find any level crossing in it for the $3 \times 3$ lattice system with the periodic boundary, as shown in the Fig.~\ref{x2d-n9}(b). 
We attribute the continuity between the Neel and collinear states to smallness of the system size, where both the states are mixed by the periodic boundaries toward both the $x$- and $y$-directions.

We examine the relation between the hole-spin bound state and a spin excited independently from the hole-spin bound state. 
For this purpose, we calculated the ground-state phase diagram for the $4 \times 4$ lattice system in the one-hole case ($N_{\rm h}=1$). 
The phase diagrams for the free and periodic boundary conditions are shown in Figs.~\ref{x2d-n16}(a) and (b). 
We find a $S=\frac{3}{2}$ phase in each figure. 
Since the number of electrons is odd, at least a spin $\frac{1}{2}$ is left unpaired. 
The profile of the $S=\frac{3}{2}$ phase is similar to that of the  $S=1$ phase of the $3\times3$ and $3\times5$ systems. 
This suggests that the $S=\frac{3}{2}$ ground state includes an independently moving $\frac{1}{2}$ spin as well as a hole-spin bound state.

\begin{figure}[t] 
\begin{center}\leavevmode
\includegraphics[width=0.9\linewidth]{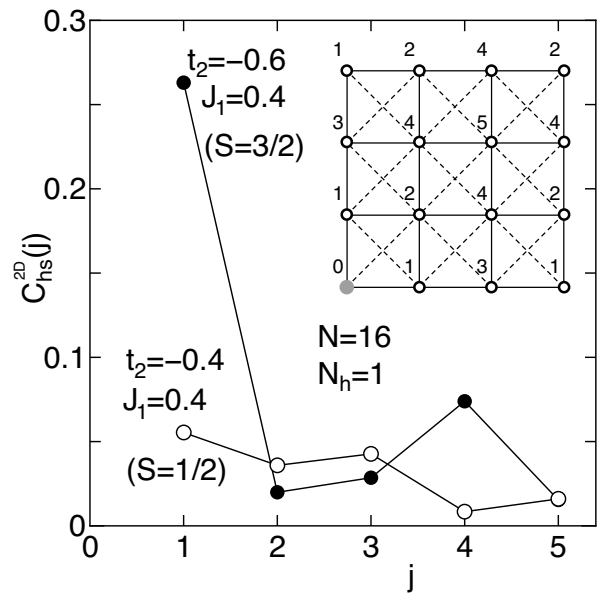}
\end{center}
\caption{
Correlation  function $C^{\rm 2D}_{\rm hs}(j)$ = $\langle \, n_{\rm h}(0, 0) \, s(i_x, i_y) \, \rangle$ between the hole and a spin for the $4 \times 4$ lattice system with the periodic boundary condition. 
Inset: The $4 \times 4$ lattice system. 
The number attached to each site represents the value of $j$ which is assigned to the distance between the hole at (0, 0) (the gray circle) and the spin at $(i_x, i_y)$ (an open circle); see text.
}
\label{Cor-n16}
\end{figure}

To examine the location of the unpaired $\frac{1}{2}$ spin relative to the hole, we calculate the 2D version of the hole-spin correlation function 
\begin{align} 
C^{\rm 2D}_{\rm hs}(j) = 
\langle \, n_{\rm h}(0, 0) \, s(i_x, i_y) \, \rangle ,
\end{align} 
where $n_{\rm h}(0, 0)$ and $s(i_x, i_y)$ are the hole number at site (0, 0) and the $z$-component of the spin at site $(i_x, i_y)$, respectively, and $j$ represents the order of the closeness between the sites (0, 0) and $(i_x, i_y)$. 
We calculated $C^{\rm 2D}_{\rm hs}(j)$ for the $4 \times 4$ lattice system by the numerical diagonalization with the periodic boundary condition. 
The results for $(J_1, t_2)$ = $(0.4, -0.6)$ in the $S=\frac{3}{2}$ phase and for $(J_1, t_2)$ = $(0.4, -0.4)$ in the upper $S=\frac{1}{2}$ phase are shown in Fig.~\ref{Cor-n16}. 

We indicate the values of $j$ for the $4\times 4$ lattice in the inset: 
as the distance from the origin is given by $r = (i_x^2 + i_y^2)^{1/2}$, 
$j$ = 1, 2, 3, 4, and 5 correspond to $r$ = 1, $\sqrt{2}$, 2, $\sqrt{5}$, and $2\sqrt{2}$, respectively. 
For $(J_1, t_2)$ = $(0.4, -0.6)$, $C^{\rm 2D}_{\rm hs}(j)$ is large for $j=1$ and the spin density is concentrated at the four sites neighboring to the hole site. 
Actually, we have $4 \times C^{\rm 2D}_{\rm hs}(1)$ = 1.052 which is close to spin 1, meaning that a triplet spin pair is formed in four sites neighboring to the hole. 
For $j$ = 2 and 3, the values of the correlation function is fairly small, suggesting that the $C^{\rm 2D}_{\rm hs}(j)$ rapidly decreases with $j$. 
This result is clearly understood if the hole and a triplet spin pair form a rigid bound state. 
We also see a small concentration at the sites with $j$ = 4. 
This suggests that the unpaired spin is repulsive to the hole-spin bound state. 
In contrast, for $(J_1, t_2)$ = $(0.4, -0.4)$ in the upper $S=\frac{1}{2}$ phase, $C^{\rm 2D}_{\rm hs}(j)$ is relatively uniform and the spin spreads over all area. 
This result reasonably means that the hole does not attract spin in the upper $S=\frac{1}{2}$ phase. 
Comparing the correlation functions for $(J_1, t_2)$ = $(0.4, -0.6)$ and $(0.4, -0.4)$, we argue that the rigid hole-spin bound state of the hole and a triplet spin pair  coexists with unpaired spin-$\frac{1}{2}$ in the $S=\frac{3}{2}$ phase. 
\begin{figure}[t] 
\begin{center}\leavevmode
\includegraphics[width=0.9\linewidth]{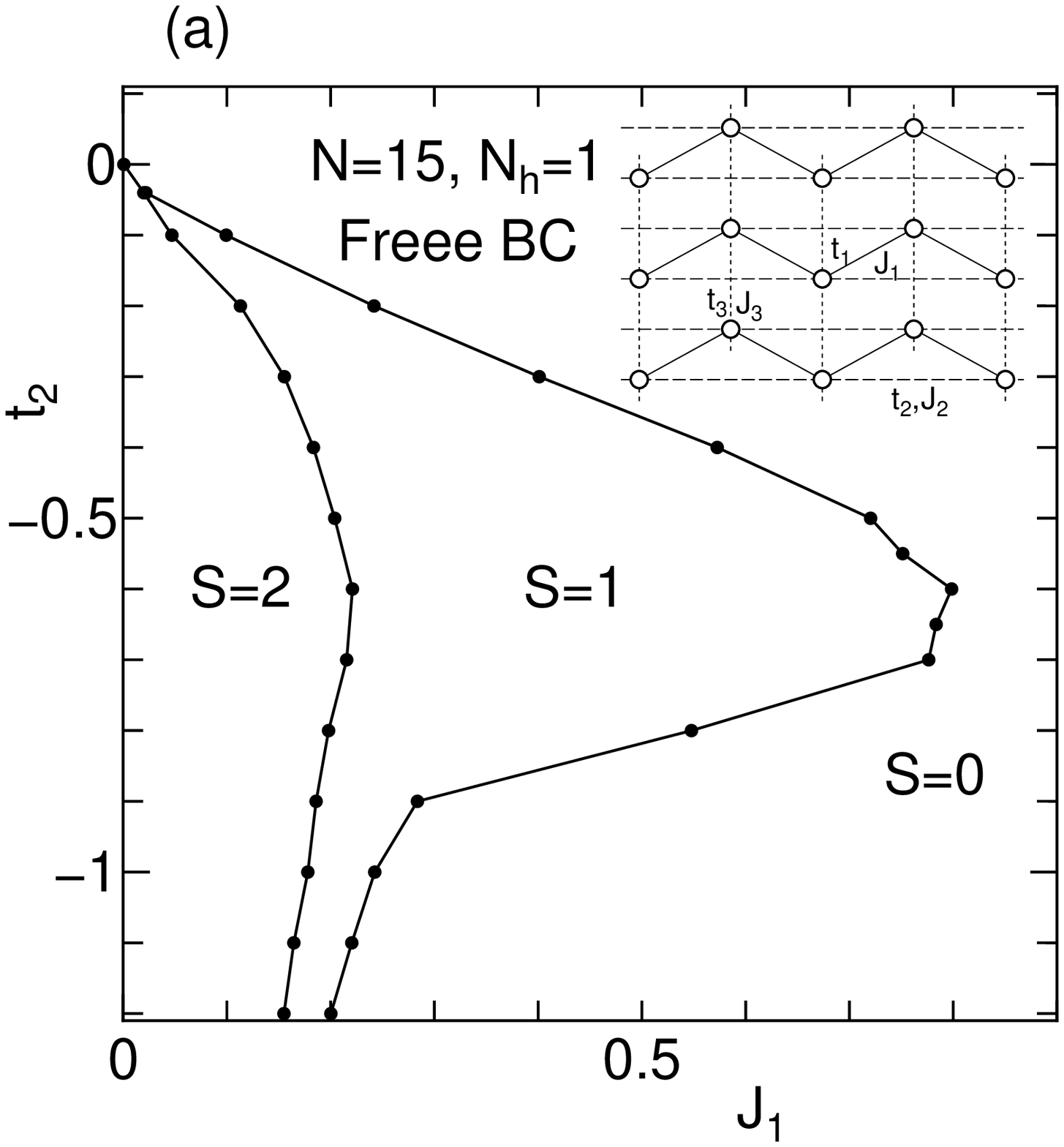}
\end{center}
\begin{center}\leavevmode
\includegraphics[width=0.9\linewidth]{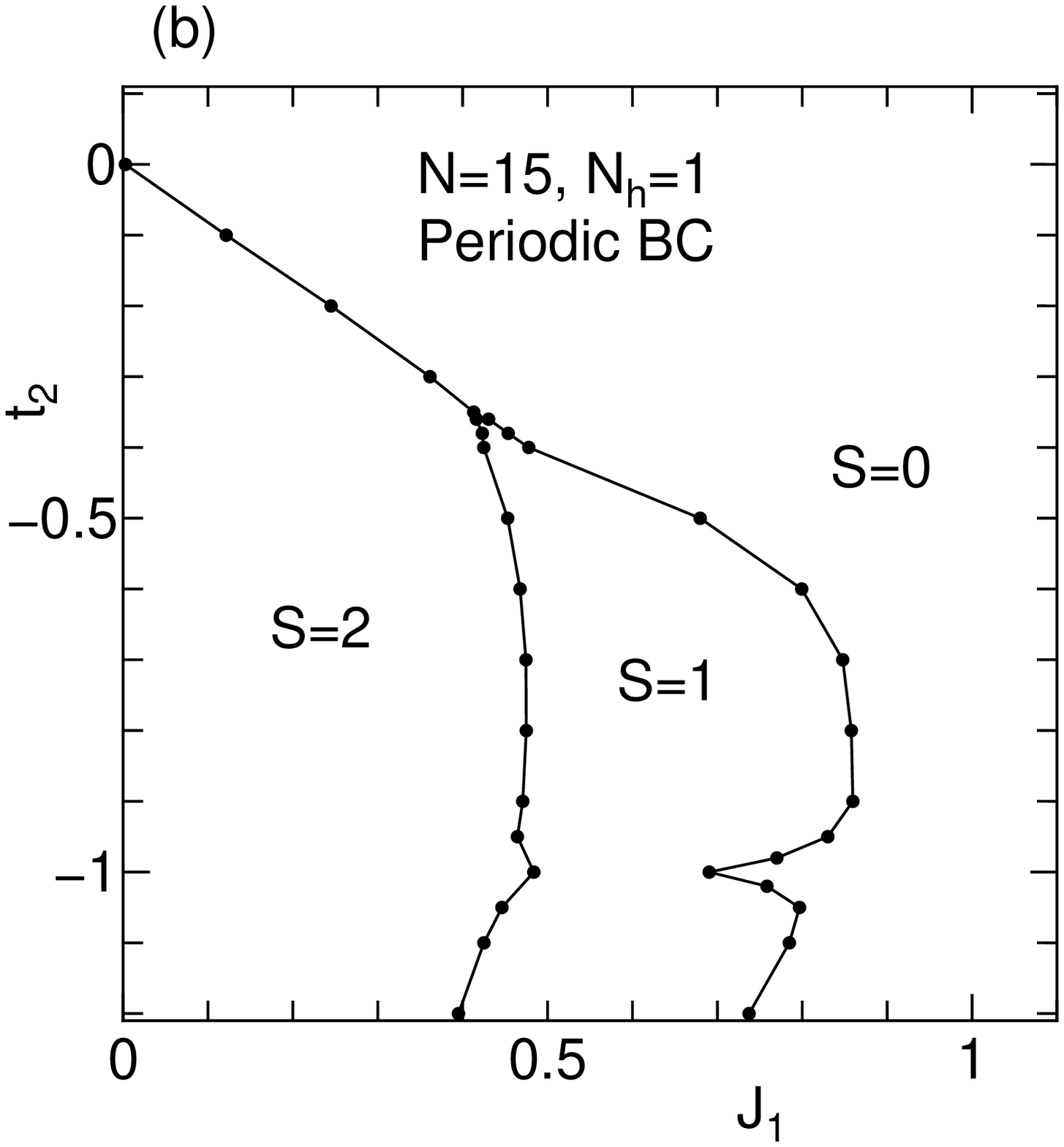}
\end{center}
\caption{Ground-state phase diagram of the quasi-1D zigzag system 
for $t_3=J_3=0.05$ (inter chain) with 
(a) the free boundary condition and 
(b)the periodic boundary condition. 
The lattice structure is shown in the inset of (a).}
\label{q1d}
\end{figure}

Finally, we briefly examine a quasi-1D system, the zigzag chain with interchain couplings. 
The lattice structure is shown in the inset of Fig.~\ref{q1d}(a), where $t_3$ and $J_3$ are the hopping and exchange energies, respectively, between zigzag chains. 
We calculated the ground-state phase diagrams for the free and periodic boundary conditions. 
The results are shown in Figs.~\ref{q1d}(a) and (b). 
We have chosen the interchain couplings as $t_3=J_3=0.05$, which is much smaller than $t_1 = 1$, the energy of a typical intrachain coupling. 
We see a $S=1$ phase around $(J_1, t_2)=(0.6, -0.6)$. 
The phase particularly for the free boundary condition is of a shape similar to those for the pure 1D and 2D cases, and also survives for the periodic boundary condition. 
The result shows that the hole-spin bound state is stable for the interchain couplings. 
Experimentally, it extends the possibility of synthesizing materials which include the present hole-spin bound state. 
We notice that there is no fully polarized ferromagnetic phase in the phase diagrams. 
The effect of Nagaoka ferromagnetism\cite{Nagaoka,Sano1987}  produced by one hole motion  cannot overcome  the effect of the antiferromagnetic interaction with small $J_3$ which is finitely fixed in the thermodynamic limit.

\section{Summary and Discussion}\label{discussion}

In summary, we investigated the bound state of a hole and a triplet spin in the 1D and 2D $t_1$-$t_2$-$J_1$-$J_2$ models  by the numerical diagonalization method. 
In the one-hole case, we confirmed that a large phase with total spin $S=1$ exists in the $J_1$-$t_2$ phase diagram for several system sizes under the free and periodic boundary conditions. 
The existence of the large $S=1$ phase indicates the existence and stability of the hole-spin bound state. 
We examined features of the hole-spin bound state by the hole-spin and density-density correlation functions. 
In the two-hole case, calculation of a specially defined spin density function showed that two holes are repulsive to each other. 
These results strongly suggest that the hole-spin bound state behaves as a tightly bounded composite particle and moves almost freely from 
the other hole-spin bound states. 

We finally aim to establish that hole-spin bound states are generally formed and behave as quasiparticles in various strongly repulsive electron systems on lattices consisting of triangle units. 
Actually, in this paper, we investigated the issue by using the 1D and 2D $t_1$-$t_2$-$J_1$-$J_2$ models as typical cases. 
Also the present arguments are based on numerical calculations with relatively small numbers of electrons and the dimensionality is less than 3 due to the limitation of the calculations. 
Hence, the present results are not general. 
However, we have obtained many substantial evidences, which are  consistent with each other, for the formation of the hole-spin bound state. 

It is plausible that the hole-spin bound state is a bosonic quasiparticle with charge $+e$ $(>0)$. 
Actually, the motion of the hole-spin bound state is equivalent to a collective back-flow motion of surrounding singlet electron pairs which are bosonic. 
Also a whole triplet electron pair as well as a whole singlet electron pair does not move with the hole for long distances due to the dense electrons with infinite on-site repulsions. 
Hence only a moving charge is $+e$ of the hole and not $-2e$ of the triplet pair near the hole. 
The hole-spin bound state consists only of a hole and a spin-1 degree of freedom in the electrons. 
According to this picture, a superconductivity may occur as the BEC of the quasiparticles of the hole-spin bound states at low temperatures. 
The supercurrent is then carried by the quasiparticles with charge $+e$ and spin 1. 
Thus we have come close to a possible exotic superconductivity of  charge $+e$. 

In the case of no hole, the BEC of massive magnons is argued. 
Oosawa et al. reported a field-induced magnetic ordering in TlCuCl$_3$, which is described by a quasi-1D zigzag spin system.\cite{Oosawa} 
The ordering is explained by the BEC of massive magnons under strong magnetic field.\cite{Nikuni,Takatsu} 
In contrast, we argued the BEC of charged particles with spin 1 in no magnetic field. 
Then the present BEC means a superconductivity. 
We expect that this type of BEC is realized in hole-doped spin-gapped materials with lattices where the triangle effect works.

\section*{ACKNOWLEDGMENTS}

The numerical computation in this work was partially carried out 
using the facilities of Information Technology Center, Nagoya University. 

\vfill



\begin{references}
\bibitem{t-tube}
The problem of triangular tube  with a hole is very interesting and we will address the detail of it  elsewhere in the future.
\end{references}
\end{document}